\documentclass[aps,reprint,superscriptaddress,nofootinbib,showpacs]{revtex4-1}
\usepackage{amsmath,latexsym,amssymb,hyperref,graphicx}
\usepackage{slashed,nicefrac}

\hypersetup{colorlinks,citecolor= nicegreen,linkcolor= nicered}
\usepackage{color}
\definecolor{nicered}{rgb}{0.7,0.1,0.1}
\definecolor{nicegreen}{rgb}{0.1,0.5,0.1}

\newcommand\eV{\text{eV}}

\newcommand\MeV{\text{MeV}}
\newcommand\GeV{\text{GeV}}
\newcommand\TeV{\text{TeV}}

\newcommand\BB{\ensuremath{0\nu2\beta}}

\newcommand\SEC[1]{\medskip\noindent{\sl\bfseries #1}}
\newcommand\PAR[1]{\medskip\noindent{\em #1}}

\begin{document}
\addtolength{\belowdisplayskip}{-.2ex}       
\addtolength{\abovedisplayskip}{-.2ex}       

\title{Neutrinoless Double Beta Decay: Low Left-Right Symmetry Scale?}

\author{Miha Nemev\v{s}ek}
\affiliation{ICTP, Trieste, Italy}
\affiliation{Jo\v zef Stefan Institute, Ljubljana, Slovenia}

\author{Fabrizio Nesti}
\affiliation{Universit\`a di L'Aquila, L'Aquila, Italy}
\affiliation{ASC, LMU M\"unchen, Germany}

\author{Goran Senjanovi\'{c}}
\affiliation{ICTP, Trieste, Italy}
\affiliation{ASC, LMU M\"unchen, Germany}

\author{Vladimir Tello}
\affiliation{SISSA, Trieste, Italy}

\date{\today}

\begin{abstract}
  \noindent 
  Experiments in progress may confirm a nonzero neutrinoless double beta decay rate in conflict with
  the cosmological upper limit on neutrino masses and thus require new physics beyond the Standard
  Model. A natural candidate is the Left-Right symmetric theory, which led originally to neutrino
  mass and the seesaw mechanism.  In the absence of cancelations of large Dirac Yukawa couplings, we
  show how such a scenario would require a low scale of Left-Right symmetry breaking roughly below
  10 TeV, tantalizingly close to the LHC reach.
\end{abstract}


\maketitle

\noindent 
If neutrinos were Majorana particles~\cite{Majorana:1937vz}, lepton number would be necessarily
violated, both in low and high energy processes. The former is exemplified by the neutrinoless
double beta decay ($0\nu2\beta$)~\cite{racahfurry}, and the latter through the Keung-Senjanovi\'c
(KS)~\cite{Keung:1983uu} production of same-sign charged leptonic pair at colliders. While \BB\ can
be mediated by Standard Model (SM) physics augmented with Majorana neutrino masses, the collider
analogue cries for new physics. A natural example of such a new physics is
provided~\cite{Keung:1983uu} in the context of Left-Right (LR) symmetric theories~\cite{lr} with
Majorana masses for heavy Right-Handed (RH) neutrinos~\cite{minkowskims}.  Present-day experiments
on neutrinoless double beta decay ~\cite{cuoregerda} are sensitive to the sub-eV region of neutrino
masses. There is even a claim of the observation of \BB\ corresponding to the neutrino masses in the
range from $0.2$ to $0.6\,\eV$~\cite{KlapdorKleingrothaus:2004wj}.

On the other hand, cosmology is setting an upper limit on the sum of neutrino
masses~\cite{Seljak:2006bg}, which keeps diminishing~\cite{Hannestad:2010yi} and makes the
contribution due to light neutrino masses potentially incompatible~\cite{Fogli:2008ig}
with~\cite{KlapdorKleingrothaus:2004wj}.  Soon, the Planck satellite will help shed light on this
important issue~\cite{hep-ph/0602058} and hopefully establish whether neutrino mass is really too
small to account for a positive finding of \BB.

In view of this, we consider seriously the possibility that new physics may be necessary~\cite{feinbergbrown}. 
Recently, we studied this issue in the context of LR symmetric theories
in~\cite{Tello:2010am} (for a review see \cite{Senjanovic:2010nq}), where we pointed out the deep connection between \BB\ and the KS same-sign leptons at colliders, in the
context of the so called type-II seesaw~\cite{typeII}. In that case, both these lepton number violating (LNV)
processes are related to lepton flavor violation, since all the RH mixing angles are predicted.

In this note we elaborate on this important connection between low and high energy experiments.  We
imagine a possible situation where \BB\ is measured and the ordinary neutrino masses can not account
for it, as described above. We focus on the minimal LR symmetric model,
without any assumption on masses and mixings in the neutrino sector.
We discuss all the mediators of the \BB\ decay and analyze the scales in this theory,
 in particular the masses of the RH gauge
boson, the RH neutrinos and the RH doubly charged component of
the triplet Higgs. As we will see below, the mass of the RH gauge boson should lie tantalizingly close
to the LHC reach, which provides further motivation for the hunt for parity restoration at colliders.

\SEC{The Model.} The minimal LR symmetric theory is based on the gauge group $\mathcal G_{LR} =
SU(2)_L \times SU(2)_R \times U(1)_{B-L}$. Fermions are LR symmetric: $q_{L,R }= (u, d)_{L,R}$ and
$\ell_{L,R} = \left( \nu, e \right)_{L,R}$ and the gauge couplings are $g_L=g_R\equiv g$.

\begin{figure*}[t]
\vspace*{-1ex}
 \centerline{%
   \includegraphics[width=0.73\columnwidth,height=.73\columnwidth]{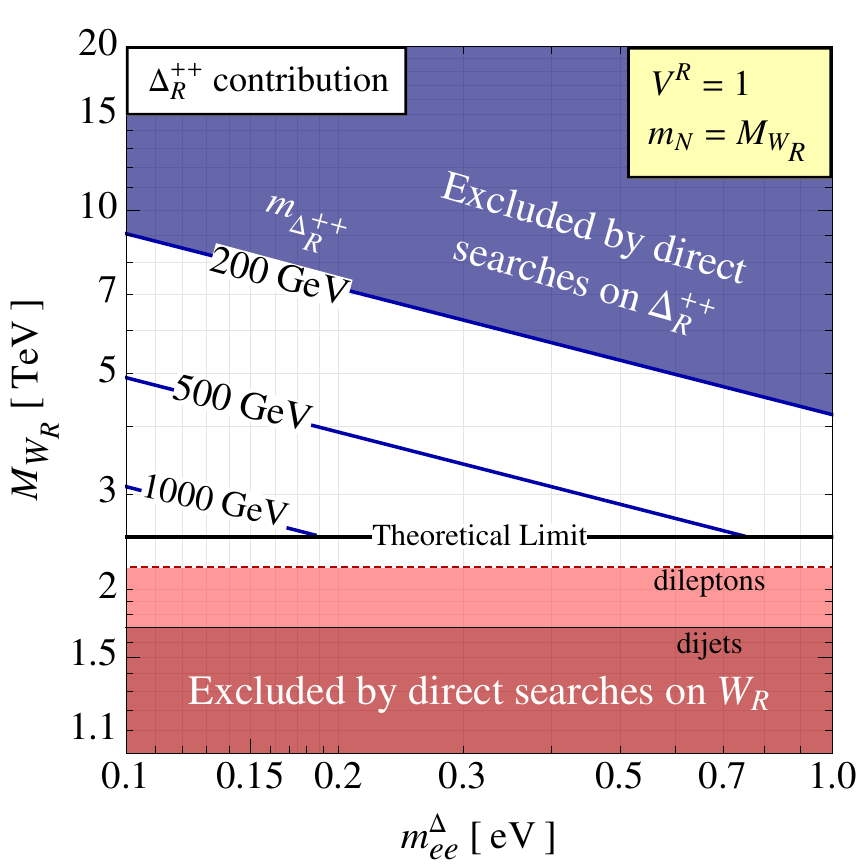}\ \ \ \ ~~~~~~
   \includegraphics[width=0.73\columnwidth,height=.73\columnwidth]{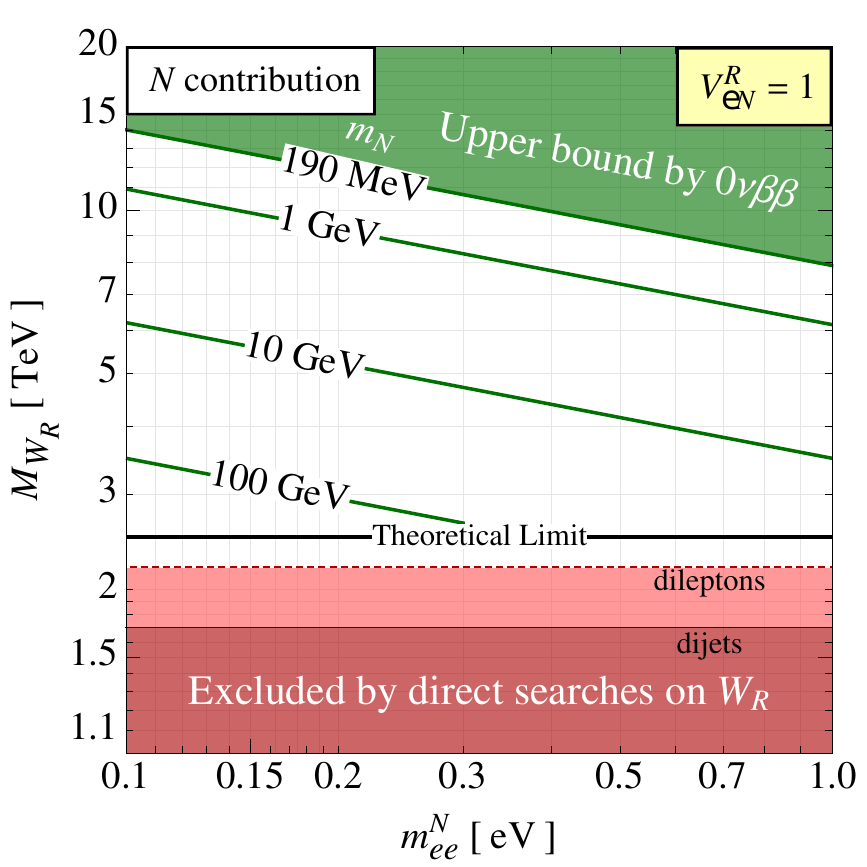}}
\vspace*{-1ex}
    \caption{Upper bound on $M_{W_R}$ from \BB\  due to the RH doubly charged scalar $\Delta^{++}_R$ (left frame) and  to the RH neutrino N (right frame).}
  \label{fig:bounds}
\end{figure*}

The Higgs sector consists~\cite{minkowskims} of the $SU(2)_{L,R}$ triplets $\Delta_{L, R} =
\left(\Delta^{++}, \Delta^+, \Delta^0 \right)_{L, R}$, $\Delta_L \in (3 , 1 ,2)$ and $\Delta_R \in
(1, 3, 2)$.  The group $\mathcal G_{LR}$ is broken down to the Standard Model (SM) gauge group by
$\langle \Delta_R \rangle \gg M_W$ and after the SM symmetry breaking, the left-handed triplet
develops a tiny $\langle \Delta_L \rangle \ll M_W$. $\langle \Delta_R \rangle$ gives masses not only
to the $W_R$ and $Z_R$ gauge bosons but also to the right-handed neutrinos.  The symmetric
Yukawa couplings of the triplets relevant for our discussion are
\begin{equation}
	{\cal L}_Y  = \frac{1}{2} \ell_L  \frac{M_{\nu_L}}{\langle \Delta_L\rangle}
	 \Delta_L \ell_L +
	\frac{1}{2} \ell_R  \frac{M_{\nu_R}}{\langle \Delta_R\rangle}  \Delta_R \ell_R+ 
\text{h.c.}\, ,
\label{eqLDelta}
\end{equation}
where $M_{\nu_L}$ and $M_{\nu_R}$ are Majorana mass matrices of light and heavy neutrinos. There is
also a bi-doublet, which contains the usual SM Higgs doublet and another, heavy flavor changing
doublet.  For a recent detailed analysis of its phenomenology and limits on its spectrum, see~\cite
{Maiezza:2010ic}. The bottom line is the lower theoretical limit $m_{W_R}\gtrsim2.5\,\TeV$. There
are also early experimental limits from LHC, still below it but rapidly catching
up~\cite{Nemevsek:2011hz,Aad:2011yg,CMS-KS}.

The bi-doublet provides the usual charged fermion and Dirac neutrino masses.  The Dirac Yukawa
couplings for neutrinos could in principle be arbitrarily small, in which case one speaks of the
so-called type II seesaw~\cite{typeII}. However, the smaller they are, the more fine-tuning one
needs, and it may be desirable to avoid this. The smaller $m_N$ is, the smaller the corresponding
Dirac Yukawa couplings have to be. The point is that neutrinos are also coupled to the heavy doublet
from the bi-doublet, with Yukawas proportional to the masses of charged leptons.  The heavy-light
doublet mixing then induces Dirac Yukawas~\cite{Deshpande:1990ip}, which leads roughly to $m_N
\gtrsim 10\, \GeV$, for $W_R$ in the 1 to 10\,\TeV\ region.

\SEC{Neutrinoless double beta decay.}  As discussed in the introduction, let us assume that \BB\
decay gets to be observed, while cosmology prevents neutrino masses from doing the job, so that new
physics must be its source. In other words, the effective mass parametrizing the \BB\ decay rate
should be appreciable: from 0.1 to 1 eV.

The simplest and most popular new physics consists of adding only RH (or sterile) neutrinos to the
SM, as to provide neutrino masses through Yukawa coupling with the ordinary LH neutrinos.  At first
glance these small couplings could not lead to dominant \BB; this is manifest in the case of a
single generation. However in the general case a mild degree of cancelation in the Yukawa couplings
could be sufficient to guarantee small neutrino masses while at the same time providing the main
source of \BB~\cite{Atre:2009rg,Tello,Ibarra:2010xw,Mitra:2011qr}. If the RH neutrinos are lighter
than about 10\,\GeV\ this possibility remains stable in perturbation theory~\cite{Mitra:2011qr}, and
thus can be considered technically natural. Since such sterile neutrinos can not be observed at
colliders, this can be considered as a conspiracy scenario in which one can see \BB\ but not
directly the new physics responsible for it.  In what follows, we refrain from considering this
phenomenological nightmare, and turn instead our attention to the contribution of the Right charged
gauge boson.

The main contribution to this process arises from the exchange of RH neutrinos ($N$) or RH doubly
charged scalar,\footnote{The joint exchange of both Left and Right gauge bosond plays a minor role.
  The same is true for the tiny $W_L$-$W_R$ mixing $\xi_{LR} < (M_W/M_{W_R})^2 \lesssim 10^{-3} \,
  \text{to} \, 10^{-4}$, and for the contribution of the bi-doublet, because of its heavy mass of at
  least 10\,\TeV~\cite{Maiezza:2010ic}. } which can be measured by the effective
mass~\cite{Tello:2010am}
\begin{equation}
	m_{ee}^{N}+m_{ee}^{\Delta} = p^2 \frac{M_W^4}{M_{W_R}^4} V_{R e j}^2
	\left[ \frac{m_{N_j}}{p^2 + m_{N_j}^2} + \frac{2 \ m_{N_j} }{ m^{2}_{\Delta_R^{++}} }  \right],
\label{eq:BB}
\end{equation}
where $p\sim190\,\MeV$ measures the neutrino virtuality and $ V_R$ is the right-handed lepton mixing matrix.



\pagebreak[3]

\SEC{Implications for the LR scale.} Let us then ask, what would the LR scenario~\cite{ms} imply for
the scale of the LR symmetry breaking, or more precisely, on the mass of the RH charged gauge boson?

As far as $\Delta^{++}_R$ is concerned, the limit $m_{\Delta_R^{++}} > (110-150) \,
\GeV$~\cite{cdfd0R++} holds, and was shown recently to be free from the uncertainties which plague
the one for $\Delta_L^{++}$~\cite{Melfo:2011nx}.  As a result, the contribution of $\Delta^{++}_R$ can
only be dominant for very heavy $N$. The interactions of $\Delta_R^{++}$ are governed by the same
$V_R$ that appears in the gauge sector and any off-diagonal entry would mediate lepton flavor
violating processes, such as $\mu \to e \bar e e$ and $\mu \to e \gamma$. The rates for all these
processes are proportional to $m_N/m_{\Delta_R^{++}}$ and in order to be safe from LFV, $V_R$ would
have to be nearly diagonal, if $\Delta_R^{++}$ were to dominate the \BB\ rate.  Still, in order to
contribute to \BB, $\Delta_R^{++}$ would have to be relatively light and therefore observable at
LHC~\cite{Perez:2008ha}.  In such case, the upper bound on $W_R$ mass is depicted in
Fig.~\ref{fig:bounds}, left frame. The lower bound on $m_{\Delta_R^{++}}$ from direct searches is
likely to increase in the near future~\cite{CMSATLASDoubly}, making this contribution less
important.

The contribution of RH neutrinos in turn can be sizable if they are light. However, it is evident
from Eq.~(\ref{eq:BB}) that the contribution decreases if their mass is lowered below $p\sim
190\,\MeV$.  Therefore, there is an upper limit on the possible contribution to \BB, and this
implies an upper bound on $M_{W_R}$. This is depicted in Fig.~\ref{fig:bounds}, right frame, where
we chose $V_{eN}^R=1$, which gives the most conservative upper bound.  Even with this extremal
choice, we see that the allowed region of $M_{W_R}$ is in the $\TeV$ region, where LHC would be able
to probe it through same-sign dileptons or leptons plus missing energy.

Notice that this result becomes even stronger if we ask of Dirac Yukawa couplings to be natural,
which amounts to $m_N \gtrsim 10\, \GeV$ as discussed above and which would guarantee observing
$W_R$ at the LHC for any value of the effective mass that requires an explanation due to new
physics.

For experimental probes of the RH neutrino, it is also important the expected range of RH neutrino
masses. In this regard, we first point out that $m_{ee}^{N}$ suffers from an ambiguity,
because it can be produced for small ($m_N \ll p$) and large ($m_N \gg p$) RH neutrino mass. Then,
we recall that the RH neutrino masses are not completely arbitrary due to cosmological
considerations, which we develop in the following section.

\pagebreak[3]

\SEC{Cosmological constraints.} If $N$'s decay after the BBN era, they end up destroying the
abundance of light elements.  This requires $\tau_N \lesssim \text{sec}$, which translates into a
lower bound on $m_N$.

Let us first consider the case of the lightest RH neutrino. It turns out that only two regimes are
allowed: the {\em heavy} regime, with $m_N \gtrsim m_\pi + m_\ell$, where $N$ decays sufficiently
fast into a charged (anti)lepton and a pion; and a {\em light} regime, with the lightest neutrino
having very low mass $m_N \lesssim$\,eV, in which case it is cosmologically stable.

\PAR{The heavy regime.} When $m_N \gtrsim m_\pi + m_\ell$, the fastest decay of $N$ is into a pion
and a lepton, which has the following decay rate:
\begin{equation}
\begin{split}
  \Gamma_{N \to \ell \pi} \!=\! \frac{G_F^2 |V_{ud}^{qR}|^2 |V_{\ell N}^R|^2 f_\pi^2 m_N^3}{8 \pi}
  \frac{M_W^4}{M_{W_R}^4} \bigl[ \left(1 - x_\ell^2 \right)^2 - 
  \\
  x_\pi^2 \left(1 + x_\ell^2 \right) \bigr] \!
  \left[ \! \left(1 \! - \! \left(x_\pi + x_\ell \right)^2 \right) \left(1\! - \! \left( x_\pi - x_\ell \right)^2 \right) \! \right]^\frac12,
\end{split}
\label{eqNDecayRate}
\end{equation}
where $x_{\pi,\ell} = m_{\pi, \ell} / m_N$, $ V^R$ is the right-handed lepton mixing matrix,
$V^{qR}$ is the analog quark one and $f_\pi = 130\,\MeV$ is the pion decay constant.  We recall that
$V_{ud}^{qR}\simeq V_{ud}^{qL}\simeq 0.97$; on the other hand, the leptonic mixing involved depends
on the mass hierarchy and on the flavor of the charged lepton into which the RH neutrino is
decaying.  As one can check from~\eqref{eqNDecayRate}, for $m_N > m_\pi+m_\ell$, the above process
guarantees that $\tau_N$ is safely shorter than a second.\footnote{There is also potentially the
  process $N \to \ell \ell \nu$, mediated through Dirac Yukawas or the LR gauge boson mixing
  $\xi_{LR}$; but in order to be fast enough it would require either large Dirac Yukawa couplings or
  large $\xi_{LR}$, both not expected and requiring some degree of cancellation in the lepton or in
  the quark sector respectively~\cite{Maiezza:2010ic}.}

We can thus summarize the resulting constraints from cosmology as
\begin{equation}
  m_N^{\rm lightest} > 140 \text{ to } 1900 \, \MeV,
\label{eq:cosmobound}
\end{equation}
where the first number is the least constraining ($\ell = e$) and the second is the most
constraining in the case when $\ell = \tau$. The actual outcome depends on the unknown mixing
angles.  Regardless of the mixing angles, below the bound in Eq.~\eqref{eq:cosmobound}, there is no
process that could make the decay of the lightest $N$ fast enough. There is then only one 
alternative, which is to go to the {\em light} regime.

\begin{figure*}[t]
\vspace*{-1ex}
 \centerline{%
    \includegraphics[width=.95\columnwidth]{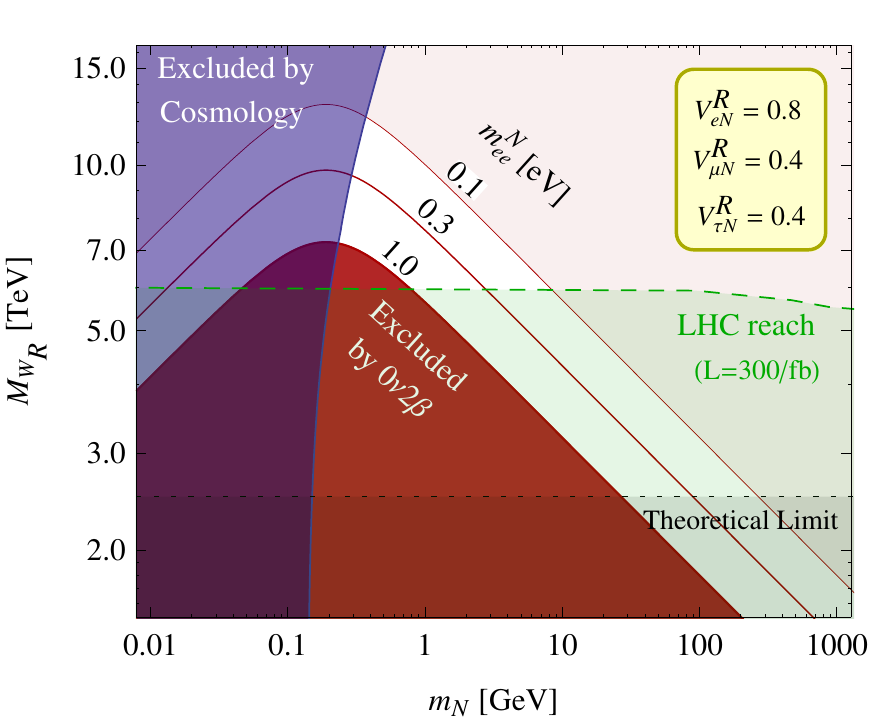}\ \ \ \ \ \ \ \ \ \ 
    \includegraphics[width=.95\columnwidth]{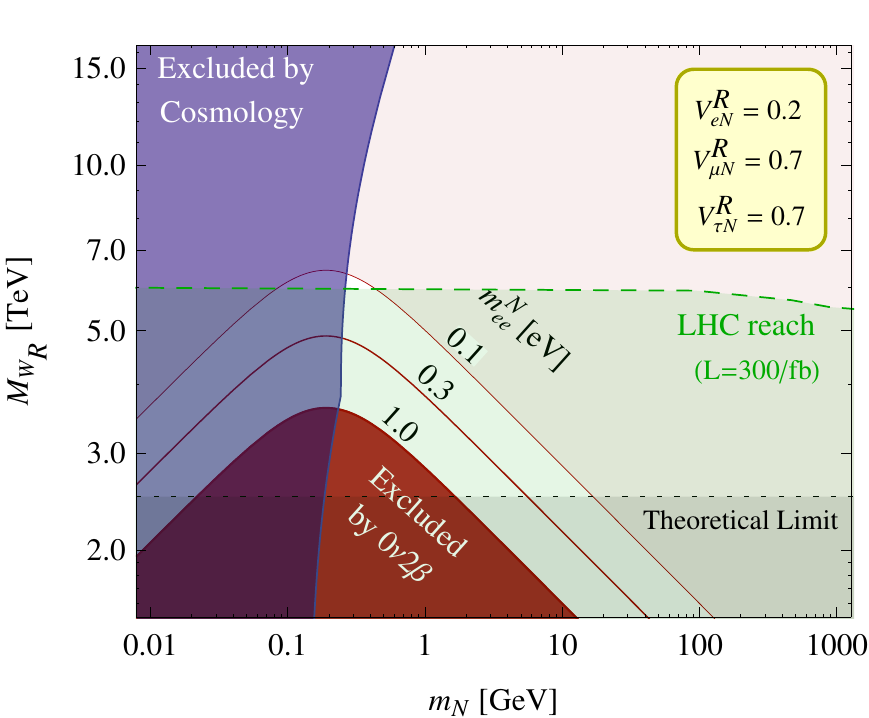}\ \ \ \ }
\vspace*{-1ex}
 \caption{Contours of $m_{ee}^N$ in the $(M_{W_R}, m_N)$ plane, illustrated for large  
    (left) and small (right) coupling to the electron $V^R_{eN}$.}
  \label{fig:contours}
\end{figure*}

\PAR{The light regime and extra species at BBN.}
Since for $m_N < 140 \,\MeV$ the lifetime becomes longer than a second, a decaying $N$ would pump
too much entropy into the universe.  The point is they decouple relativistically at the temperature
\begin{equation}
  T_D^N= T_D^\nu \left(\frac{M_{W_R}}{M_{W}}\right)^{\frac43},
\end{equation}
where $T_D^\nu\simeq1\,\MeV$ is the neutrino decoupling temperature. Therefore, for a representative
value of $M_{W_R}\sim 5\,\TeV$,
\begin{equation}
  T_D^N\simeq 250\,\MeV\,.
\end{equation}
Then, since between $T_D^N$ and 1\,\MeV\ only muons and pions decouple, at BBN $N$'s are almost
equally abundant as light neutrinos. The only way out would be to make $N$ stable and to avoid the
over-closure of the universe, lighter than about \eV~\cite{Seljak:2006bg, Fogli:2008ig,
  arXiv:1010.3968}.\footnote{It is worth mentioning a potential loophole in this argument. The
  stable neutrino could in principle be heavier, if its abundance were diluted by decays of other
  species, before the onset of BBN. This scenario was studied in~\cite{Bezrukov:2009th}, where the
  authors concluded that $M_{W_R}\gtrsim 10\,\TeV$ for this to work. In this case, the second RH $N$
  has to be heavier than a few GeV and thus \BB\ decay can never be dominated by new physics.} It is
easy to check that for such a light neutrino, its lifetime is much longer than the age of the
universe. As a result, we are in a scenario where extra species are contributing to BBN.  Actually,
this situation seems to be preferred and a recent study suggests~\cite{Hamann:2010bk} that four
light neutrinos give the best fit to cosmological data, while five is disfavoured and six is
basically excluded.

There is an important astrophysical bound on the LR theory with a light $N$ coupled to electrons
coming from supernovae cooling~\cite{SCIPP-87/107, MasterYue}. For the minimal LR model with equal left and right
quark mixing angles,~\cite{UMPP-88-205} finds a lower limit $M_{W_R} \gtrsim 23 \, \text{TeV}$ for a
small LR gauge boson mixing assumed here.\footnote{Reference~\cite{UMPP-88-205} also finds a light
  $W_R$ region of about 500 GeV, today excluded by experiments, in the case of the minimal model.}
However, this is strongly subject to the right-handed lepton flavor mixing matrix $V_{\ell
  N}^R$. The $W_R$ can be as light as one wishes, as long as the light $N$ is sufficiently decoupled
from the electron; the true limit reads
\begin{equation}
	V_{eN}^R \left( \frac{2.3 \, \text{TeV}}{M_{W_R}} \right)^2 < 0.01.
\end{equation}
This is a rather strong constraint on light RH neutrinos and $W_R$ in the LHC reach, the regime of
great interest. For example, the popular type II seesaw scenario where the left and right leptonic
mixing angles are the same is ruled out in the case of normal hierarchy, while the inverse hierarchy
could work only with a very small $\theta_{13}$, disfavored by recent measurements.

It is also interesting to note that while ${N}^{\rm lightest}$ escapes the cosmological bounds
\eqref{eq:cosmobound} by being very light, the heavier RH neutrinos still have to be heavy enough
to be BBN safe.  The relevant process here is $N^{\rm heavy}$ decaying to the lightest $N$ and two
charged leptons via $W_R$, whose decay width is
\begin{equation}
  \Gamma_{N_i \to N_j\ell \ell'} \! = \! 2 \Gamma_\mu \! 
  \left(\! \frac{M_W}{M_{W_R}} \! \right)^4 \! \left( \! \frac{m_N}{m_\mu} \!\right)^5 \!
  \left| \! V_{\ell i}^R V_{\ell' j}^{R*} + V_{\ell j}^R V_{\ell' i}^{R*} \! \right|^2 \!.
\label{eqNmidDecayRate}
\end{equation}
This gives a bound on the heavier neutrino mass 
\begin{equation}
  \label{eq:cosmobound2}
  m_{N}^{\rm heavy} \gtrsim 100 \, \MeV \left(\frac{M_{W_R}}{2.26 \, \TeV} \right)^{4/5} \text{ to } (m_\mu + m_\tau),
\end{equation}
which also depends on the mixings; e.g.\ the higher end refers to the $\mu\tau$ channel.

\SEC{Interplay with LHC.}
%
%
Let us start with the canonical, SM-like scenario at low energies: only three light neutrinos below
eV.  In such a case, the limit on the mass of the lightest $N$ given in~\eqref{eq:cosmobound}
applies. The cosmological limit is actually dependent on the mass of $W_R$, and its interplay with
\BB\ is shown in Fig.~\ref{fig:contours}, illustrated in the case of large and small electron mixing.

The first result to be noticed is that of the two solutions with smaller and larger $m_N$, the
cosmological considerations allow only the heavy $N$ solution. This is also welcome in view of the
possibility to detect RH neutrinos at LHC, since a light $N$ would decay out of the detector and
only manifest itself as missing energy.  The second observation is that the values of $m_N$ and
$M_{W_R}$ required for \BB\ lie in the region where LHC will be able to probe them through same-sign
dileptons or lepton plus missing energy~\cite{Ferrari:2000sp}.  

\medskip

In the scenario where one or more $N$'s in the light regime play the role of extra light species at
BBN, the heavier $N$ still have to satisfy the cosmological constraint given
in~\eqref{eq:cosmobound2}. As a result, a similar situation regarding $W_R$ emerges.  If
$n_{\text{BBN}} = 4$, basically the same analysis goes through as above. On the other hand, if
$n_{\text{BBN}} = 5$ turned out to be the preferred option, one would have a clear
prediction. Namely, since the two light $N$'s would have to decouple from the electron due to the SN
constraint, the heavy $N$ would couple maximally and the left diagram in the Fig.~\ref{fig:contours}
would apply.


\SEC{Discussion and outlook.} 
Cosmology keeps lowering the limit on the sum of neutrino masses, which may soon come to clash with
possible evidence of \BB. If so, new physics would be mandatory. This new physics could just be RH
neutrinos, assuming cancelations among Dirac Yukawa couplings. In this work we try to avoid this
scenario, which leads us to consider the contribution of new RH gauge interactions.  In the context
of the LR symmetric model, we show how the mass of the RH charged gauge boson has to be less than
about 10 TeV, thus very close to the 14 TeV LHC reach with a high luminosity.  Another important
conclusion which emerges from cosmological considerations is that only the heavy ($m_N \gg 100 \,
\MeV$) contribution to \BB\ is allowed, therefore a measurement of the double beta process would
relate unambiguously the masses of $N$ and $W_R$.

The situation is particularly exciting in view of the simultaneous time scale for low energy
experiments such as \BB\ and LFV, the expected results from the Planck satellite and of course at
high energies from the LHC.

\SEC{Acknowledgments.}  We are grateful to Alejandra Melfo, Francesco Vissani and Yue Zhang for
important discussions and comments. F.N. and G.S. are grateful to Alexander von Humboldt Foundation
for the hospitality at the LMU during the last stages of this work.  The work of G.S. is supported
in part by the EU grant UNILHC- Grant Agreement PITN-GA-2009-237920.


\end{document}